\documentclass[twocolumn,showpacs,preprintnumbers,amsmath,amssymb,prl]{revtex4}

\usepackage{graphicx}
\usepackage{dcolumn}
\usepackage{bm}

\usepackage{wasysym}

\begin{document}

\title{\textbf{FIRST OBSERVATION OF FESHBACH RESONANCES AT VERY LOW MAGNETIC FIELD IN A $^{133}$CS FOUNTAIN.}}

\author{H. Marion}
    \altaffiliation[Present address: ]{Institut d'Optique
    Théorique et Appliqu\'ee. Bat 503, Centre Scientifique d'Orsay, 91403, Orsay, France}
\author{S. Bize}
\author{L. Cacciapuoti}
    \altaffiliation[Present address: ]{Dipartimento di Fisica, Universita degli studi di Fireze, via Sansone 50019 Steso Firentino (FI), Italy}
\author{D. Chambon}
\author{F. Pereira dos Santos}
\author{G. Santarelli}
\author{P. Wolf}
    \altaffiliation[Present address: ]{Bureau International des Poids et Mesures, Pavillon de
    Breteuil,
92312, S\`evres cedex, France }
\author{A. Clairon}
    \affiliation{BNM-SYRTE, Observatoire de Paris, 61 Avenue de l'Observatoire, 75014 Paris, France. }
\author{A. Luiten}
\author{M. Tobar}
    \affiliation{The University of Western Australia, School of Physics, 35 Stirling Hwy, Crawley, Western Australia.}
\author{S. Kokkelmans}
    \altaffiliation[Present address: ]{Eindhoven University of Technology, P.O.~Box~513, 5600~MB  Eindhoven, The Netherlands}
\author{C. Salomon}
    \affiliation{Laboratoire Kastler Brossel, ENS, 24 rue Lhomond, 75005 Paris, France.}

\begin{abstract}
For a long time, one of the main limitations of cesium atomic
fountains has been the cold collision frequency shift
\cite{Chin01,Bijlsma94,Gibble93,Ghezali96, Wynands}. By using a
method based on a transfer of population by adiabatic passage (AP)
\cite{Messiah,Loy74,Pereira02} allowing to prepare cold atomic
samples with a well defined ratio of atomic density as well as
atom number we have a better measurement of this effect. By
improving the method we found out some unexpected properties of
cold collisions, confirming that $^{133}$Cs is a rich atom to
study cold collisions. Those results lead to fine comparisons with
cold collision theory and may constraint some parameters of
cesium.\\


\end{abstract}


\maketitle




Frequency standards are used in many industrial (Global
Positioning System, navigation) and scientific pursues
(fundamental physic tests \cite{Marion03,Bize03,TheseHarold}).
Primary standards are today based on cesium and are contributing
to the realization of the second. The most accurate are using cold
atoms to allow long interaction time. In cold atomic gases
collisions play an important role by shifting the clock frequency.
Thus, it is crucial to control and to study accurately cold
collisions for clock uncertainty budget. Nevertheless, there is a
tradeoff between the clock stability and its accuracy. In this
paper we will describe improvement of the a technic based on fast
AP that allows to reconcile those two goals. For high density
atomic sample, such as in Bose-Einstein condensates (BEC)
\cite{Anderson95}, manifestations of collision such as molecular
Feshbach resonances \cite{Feshbach,Tiesinga99} are observed. We
found out, in our fountain, such phenomenon at very low magnetic
field and density. Theoretical tools will be remind and a
numerical simulation of some of our data will be presented.


\section{The double fountain: recent improvements and performances.}
\subsection{Atom loading improvement.}

\begin{figure}
  \includegraphics[width=8cm]{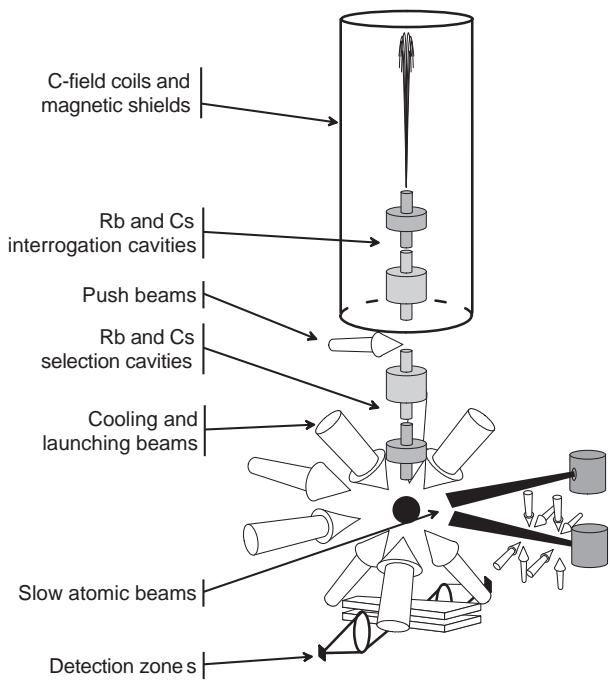}\\
  \caption{Schematic of the double fountain both base on $^{87}$Rb and $^{133}$Cs, so
that there is two pairs of cavity finely tuned on hyperfine atomic
frequency of corresponding atoms and two ovens. On the cesium
part, a transversal collimation of the atomic beam has been
improved in addition of the present chirp cooling. This consists
in a 2D recycling light, pair of laser beam, used in a ``zigzag"
way. This allows to load around $10^{9}$ atoms in  200 ms in an
optical molasses.}\label{fig:FontaineRbCs5}
\end{figure}

 At Paris observatory, there is a set of three
atomic fountains, using cesium. In particular, our one is a double
fountain \cite{Bize01, TheseHarold, Wynands} able to work
alternatively with rubidium and cesium. In the last year, on the
cesium part, we improved the loading rate by supplementing a
transverse collimation of the atomic beam in addition to the
present chirp cooling (see FIG.~\ref{fig:FontaineRbCs5}). It
consists in a 2D optical molasses with two laser beams, recycling
the light in a ``zigzag" way \cite{shimizu90,rasel99}. This allows
us to cool $10^9$ atoms in 200~ms in an optical molasses. The 2D
molasses increase the loading rate by a factor of $\sim10$.
Regular parameters of the double fountain are a magnetic field in
the interrogation zone of 2~mG, a loading duration of $\sim$ 200
ms that allows to load $10^{9}$ atoms within a fountain cycle of
$\sim$ 1.2 s. The launch velocity is $\sim$ 4 m.s$^{-1}$ that
imply a launch height of $\sim$ 0.8 m.

\subsection{Frequency stability.}

A sapphire cryogenic oscillator (SCO) \cite{luiten95} from the
University of Western Australia is now used as a filter for one of
our H-maser. The SCO is an experiment by itself \cite{Wolf03}, but
is very reliable and can be used routinely to drive the fountain.
The SCO is weakly phase-locked on the H-maser with a time constant
of about an hour. Therefore, the resulting signal reflects the
good short term stability of the SCO while in long time scales it
reproduce the characteristics of the H-maser. The SCO delivers a
signal of about 12~GHz.

At BNM-SYRTE most of the microwave synthesizer use a 100~MHz
signal to generate the 9.2~GHz necessary to interrogate the atoms.
Therefore, a down-converter has been developed to transform the
SCO signal into a 100~MHz one. An evaluation has shown that the
fountain was limited by the phase noise of the microwave setup.
Henceforth, a new microwave synthesizer \cite{chambon04} able to
down-convert directly the 12~GHz to 9.2~GHz has been built in
order to minimize the phase noise. It achieves a frequency
stability of $\sim 3\ 10^{-15}$ at 1s. In order to take benefits
of its excellent short term stability ($\sim5\ 10^{-16}$ at 1s) a
large number of trapped atoms is required \cite{Santarelli99}.
With $\sim10^{7}$ detected atoms we were able to reach $1.6\
10^{-14}$ at 1~s, the best frequency stability for a primary
standard to date. With this performances a resolution of
$10^{-16}$ is achievable in 6~hours.

\begin{figure}
  \includegraphics[width=8cm]{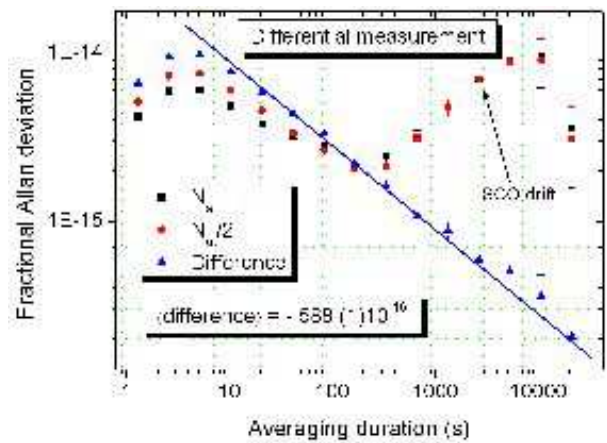}\\
  \caption{Shows the Allan deviation of the fountain driven by the SCO. This
is a differential measurement of the collisional shift, this means
that a full atom number configuration
$(${\footnotesize$\blacksquare$}$)$ is alternated with a half atom
number configuration ({\Large$\bullet$}) each 60 fountain cycles.
The $(${\footnotesize$\blacksquare$}$)$ represents the fractional
frequency deviation of the corrections applied to the
interrogating microwave generated with the SCO signal to stay at
resonance with the atoms. This is performed with the maximum atom
number which allows the $1.6\ 10^{-14}$ at 1~s stability. The
Allan deviation of the difference ({\large$\blacktriangle$})
averages well for all times and has also a good stability. This
leads to a good statistical evaluation of the mean difference in a
very reasonable time.}\label{fig:VarDiff}
\end{figure}

FIG.~\ref{fig:VarDiff} shows the Allan deviation of the fountain
driven by the SCO. This is a differential measurement
\cite{Sortais00}, this means that a full atom number configuration
$(${\footnotesize$\blacksquare$}$)$ is alternated with a half atom
number configuration ({\Large$\bullet$}) each 60 fountain cycles.
The $(${\footnotesize$\blacksquare$}$)$ represents the fractional
frequency deviation of the corrections applied to the
interrogating microwave generated with the SCO signal to stay at
resonance with the atoms. This is performed with the maximum atom
number ($N_{at}$) which allows the $1.6\ 10^{-14}$ at 1s
stability. On short term (10~s to 100~s)
 the allan variance is dominated by the Cs fountain noise. After 200~s, the drift of the SCO is observed. The Allan
deviation of the frequency difference between low ($N_{at}/2$) and
high($N_{at}$) density ({\large$\blacktriangle$}) averages as
white frequency noise for all times and reaches $\sim2\ 10^{-16}$
for a 20000~s measurement. This leads to a good statistical
evaluation of the mean difference in very reasonable time. The SCO
drift is efficiently rejected and atom number dependent effects
(i.e. cavity pulling \cite{Bizeieee01} and collisional shift) are
measured with a statistical uncertainty of $10^{-16}$ in one day.
This behavior indicates that determination of the collisional
frequency shift is only limited by the frequency stability of the
fountain.

\section{Accurate control of cold collision frequency shift.}

In previous work \cite{Pereira02, TheseHarold}, we demonstrated
the possibility to control and evaluate the cold collision shift
at the percent level by using the adiabatic passage method. Cold
collisions represent the main systematic effect that shifts the
frequency and restricts the accuracy in cesium fountains. At that
time, our collisional shift for full density, was around
$10^{-14}$. Therefore, the corresponding uncertainty was near
$10^{-16}$.

\subsection{Adiabatic passage performances.}

To achieve a stability around $\sim 2\ 10^{-14}$ at 1~s it is
necessary to load a high number of atoms. For example the
collisional shift corresponding to $N_{at}$ of the
($\blacksquare$) curve implies a frequency shift of about
$\sim1.2\ 10^{-13}$. Thus, an AP at the percent level is no longer
enough to stay at the $10^{-16}$ level of uncertainty on this
effect. To perform AP the amplitude of the microwave into the
selection cavity has to be modulated and its frequency has to be
swept. It is critical that the synthesizer used to chirp the
selection microwave stops the frequency sweep at resonance for the
half AP. Our model (Marconi 2030) was not accurate enough and
failed to be reproducible. It has been replaced by a DDS more
easily frequency controllable. According to \cite{Pereira02,
TheseHarold}, one observable for AP evaluation is the ratio of
atom number at full and half density. Each measurement is daily
analyzed using this criterion to verify that the AP method is
properly working.

\begin{figure}
  \includegraphics[width=8cm]{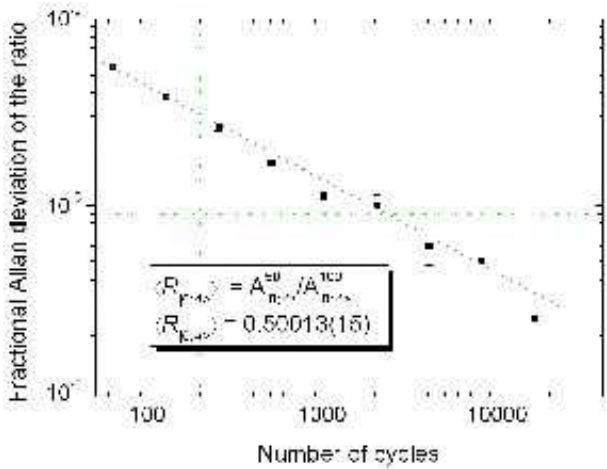}\\
  \caption{Allan deviation of the
ratio of atom numbers in the $|4; 0\rangle$ obtained by
alternating each 60 fountain cycle a full atom number
configuration over half atom number configuration on the $|4;
0\rangle$ state as a function of cycle number. The Allan deviation
of the ratio averages as white noise and reaches $\sim 10^{-3}$
resolution after 1 hour. It leads to evaluate AP at the $10^{-3}$
level.}\label{fig:VarRatio}
\end{figure}

FIG.~\ref{fig:VarRatio} shows the Allan deviation of the ratio of
atom numbers in the $|4; 0\rangle$ obtained by alternating each 60
fountain cycle a full atom number configuration over half atom
number configuration as a function of cycle number. The Allan
deviation of the ratio averages as white noise and reaches $\sim
10^{-3}$ resolution after 1 hour. It leads to evaluate AP at the
$10^{-3}$ level such as:

\begin{equation*}
\frac{A^{(50)}_{|4; 0\rangle}}{A^{(100)}_{|4;
0\rangle}}=\frac{\tilde{n}^{(50)}_{|4;
0\rangle}}{\tilde{n}^{(100)}_{|4;
0\rangle}}=\frac{1}{2}\times(1+1\ 10^{-3})
\end{equation*}

Where $\tilde{n}^{(100)}_{|4; 0\rangle}$ ($\tilde{n}^{(50)}_{|4;
0\rangle}$) is the density of the $|4; 0\rangle$ atoms for full
atom number (half atom number) and $A^{(100)}_{|4; 0\rangle}$
($A^{(50)}_{|4; 0\rangle}$) is the detected atom number on the
$|4; 0\rangle$ for full atom number (half atom number).

The deviation measured at the $10^{-3}$ level could be induced by
a non linearity in the detection. This will be investigated in a
future work.

Similarly, we also check routinely the same ratio on the F $= 3$
level. Here, we find a larger deviation:

\begin{equation*}
\frac{A^{(50)}_{|3; 0\rangle}}{A^{(100)}_{|3;
0\rangle}}=\frac{\tilde{n}^{(50)}_{|3;
0\rangle}}{\tilde{n}^{(100)}_{|3;
0\rangle}}=\frac{1}{2}\times(1+3\ 10^{-3})
\end{equation*}

The departure of $3\ 10^{-3}$ from the $1/2$ ratio is due to
residual populations in the $|3; m_{\text{F}}\rangle$ states. A
fraction of the atoms is coming from the push beam (see
FIG.~\ref{fig:FontaineRbCs5}, i.e. a laser beam that finalize to
the selection process) which is de-pumping atoms from the $|4;
m_{\text{F}}\rangle$ states to the $|3; m_{\text{F}}\rangle$
states. The remanning fraction seem to be launched in the
$\text{F}=3$ state. Attempts to remove these atoms with extra
repumper beams where partially unsuccessful. At this point, there
is nothing much to do to prevent it but to look at the
contribution of those spurious atoms on the clock shift.

\subsection{Towards an adiabatic passage at the $10^{-3}$ level.}

An accurate control of the collisional shift is possible only if
the contribution of the 0.3 \% atoms populating states different
from the clock states is correctly evaluated. To measure this, we
intentionally populated the atomic cloud using AP with
$|3;m_{\text{F}}\neq0\rangle$ states in addition of the
$|3;0\rangle$ clock state. An additional microwave synthesizer
tuned on a
$|4;m_{\text{F}}\neq0\rangle\rightarrow|3;m_{\text{F}}\neq0\rangle$
transition is combined in the selection cavity with the regular
selection synthesizer. A magnetic field is applied during the
selection by AP in order to prevent parasitic excitations between
zeeman sub-levels. To select, for example $m_{\text{F}}=1$ state,
one has to detune from resonance the extra synthesizer of the
corresponding first order Zeeman shift value.

The measurement method alternate three different fountain
configurations. The first one is a full AP selection of the $|3;
0\rangle$ state, the second is a half AP selection of the $|3;
0\rangle$ state. The first configurations give the collisional
shift due to the clock states. The third one is a full AP
selection of the $|3; 0\rangle$ state plus full AP selection of
one of the $|3; m_{\text{F}}\neq0\rangle$ states. The last
configuration combined with the two others lead to a measurement
the collisional shift due to the extra population. The respective
frequency shifts are:

\begin{eqnarray}
\nonumber  \Delta\nu_{1}/2\pi &=& \tilde{n}_{00}K_{00} \\
\nonumber  \Delta\nu_{2}/2\pi &=& \frac{\tilde{n}_{00}}{2}K_{00}\\
\nonumber  \Delta\nu_{3}/2\pi &=&
\tilde{n}_{00}K_{00}+\tilde{n}_{m_{\text{F}}m_{\text{F}}}K_{m_{\text{F}}m_{\text{F}}}
\end{eqnarray}

Where $\tilde{n}_{00}$ ($\tilde{n}_{m_{\text{F}}m_{\text{F}}}$)
represents the atom density of the clock states ($m_{\text{F}}$
states), and $K_{00}$ ($K_{m_{\text{F}}m_{\text{F}}}$) is the cold
collisional shift coefficient resulting of the extrapolated to
zero density for the clock states ($m_{\text{F}}$ states).

Under regular operating conditions, we evaluate the contribution
to the clock shift due to the collisional shift of the extra $|3;
m_{\text{F}}\neq0\rangle$ atoms per detected atom
($\nu_{m_{\text{F}}m_{\text{F}}}$ in Hz.(detected atoms)$^{-1}$)
normalized to the collisional shift of the clock states $|3;
0\rangle$ atoms per detected atoms ($\nu_{00}$) as a function of
the $m_{\text{F}}$ value. This ratio ($R_{0m_{\text{F}}}$) can be
expressed as:

\begin{eqnarray}
\nonumber  R_{0m_{\text{F}}} &=&  \frac{\nu_{m_{\text{F}}m_{\text{F}}}}{\nu_{00}}\\
\nonumber  &=&
\frac{\Delta\nu_{3}-\Delta\nu_{1}}{2(\Delta\nu_{1}-\Delta\nu_{2})}\frac{{N_{00}}}{N_{m_{\text{F}}m_{\text{F}}}}
\end{eqnarray}

Were $N_{00}$ ($N_{m_{\text{F}}m_{\text{F}}}$) is the detected
atoms number for the clock states ($m_{\text{F}}$ state) and can
be expressed as $N_{00}=A_{|3; 0\rangle}+A_{|4; 0\rangle}$. A
verification that $\nu_{00}$ is field independent has been
performed so that it could be used as a reference. It is important
to ensure that the density is the same whatever the could is made
of. Thus a verification that the spatial distribution of the
different Zeeman sub-levels is homogenous into the cloud has been
performed. By selecting one $m_{\text{F}}$ at the time and pulsing
the Ramsey cavity with different durations, we access at different
cloud slice. It shows that the peak density are equal to less than
7~\%.

\begin{figure}
  \includegraphics[width=8cm]{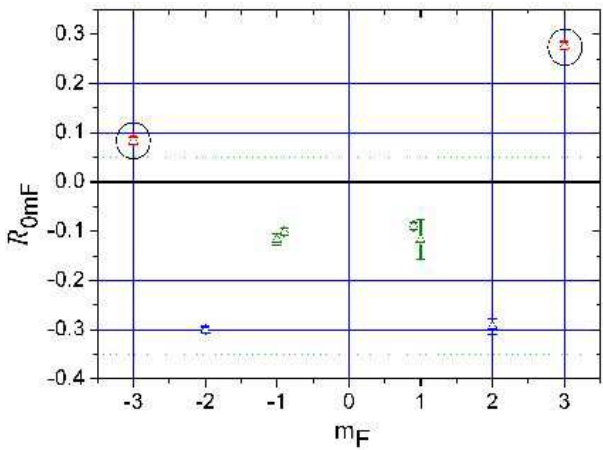}\\
  \caption{displays $R_{0m_{\text{F}}}$ as a
function of the $m_{\text{F}}$ value for regular clock conditions.
The contribution to the cold collision shift of the
$|m_{\text{F}}|= 1;~2$ are negative and equal for the same
absolute values. At most the contribution of spurious $|3;
m_{\text{F}}\neq0\rangle$ is 1/3 of the contribution of clock
states. The contribution of the $|m_{\text{F}}|=3$ is
different.}\label{fig:ShiftMfChampNormal}
\end{figure}

FIG.~\ref{fig:ShiftMfChampNormal} displays $R_{0m_{\text{F}}}$ as
a function of the $m_{\text{F}}$ value for regular conditions
($B\simeq2~mG$). The contribution to the cold collision shift of
the $|m_{\text{F}}|= 1;~2$ are negative and equal for the same
absolute values. At most the contribution of spurious $|3;
m_{\text{F}}\neq0\rangle$ is 1/3 of the contribution of clock
states. Thus, the collisional shift is controlled at the $10^{-3}$
level. Surprisingly the contribution of the $|m_{\text{F}}|=3$ is
different, which was unpredicted at such low magnetic field. This
observation is a clear indication of the presence of Feshbach
resonances.


\section{Feshbach resonances.}

\subsection{Experimental results.}

\begin{figure}
  \includegraphics[width=8cm]{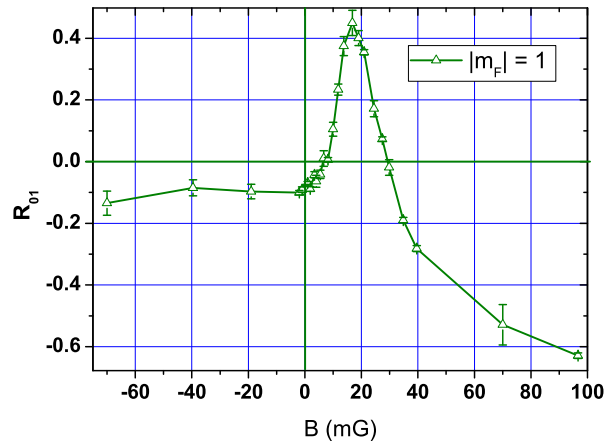}\\
  \caption{Ratio $R_{01}$ as a function of the magnetic field. It represents a Feshbach resonance for $|m_{\text{F}}|=1$.}\label{fig:Reso1}
\end{figure}

\begin{figure}
  \includegraphics[width=8cm]{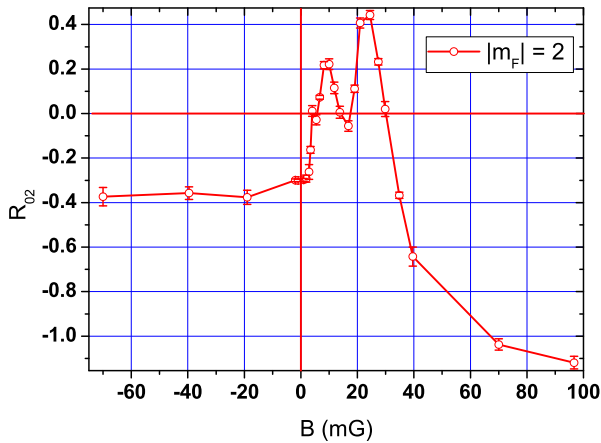}\\
  \caption{Ratio $R_{02}$ as a function of the magnetic field. It represents a Feshbach resonance for
   $|m_{\text{F}}|=2$. This is obviously a multiple resonance.}\label{fig:Reso2}
\end{figure}

\begin{figure}
  \includegraphics[width=8cm]{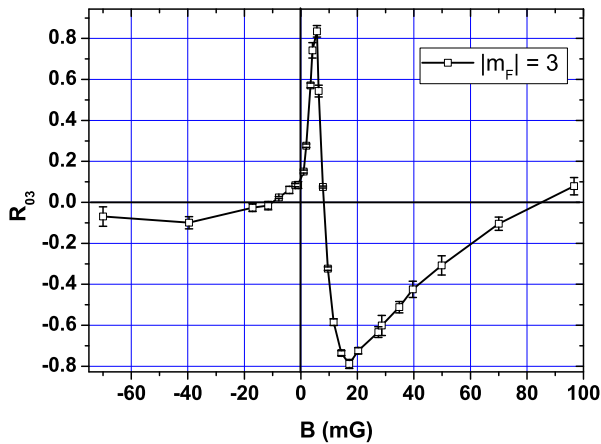}\\
  \caption{Ratio $R_{03}$ as a function of the magnetic field. It represents a Feshbach resonance for
   $|m_{\text{F}}|=3$. This resonance has a more simple shape. Assuming that it is a simple resonance we gone
to simulate the scattering process.}\label{fig:Reso3}
\end{figure}

We thus looked for the ratio ($R_{0m_{\text{F}}}$), for each
absolute value of the $m_{\text{F}}$ as a function of magnetic
field. We found out three Feshbach resonances
(FIGs.~\ref{fig:Reso1}, \ref{fig:Reso2} and \ref{fig:Reso3})
deeply field dependent. It gives a direct access to the amplitude
and the width of each resonance. The field is very well controlled
so that there is no way that error bars could mix up the shapes.
For some values of the magnetic field the frequency shift due to
$|3; m_{\text{F}}\rangle\neq0$ population can be as large as the
standard clock shift. Let's remind that our regular magnetic field
is 2~mG. One has to be very cautious by choosing a value of
magnetic field in an experiment using cold $^{133}$Cs. The
$|m_{\text{F}}|=2$ resonance is obviously a multiple resonance,
but the $|m_{\text{F}}|=1$ and specially $|m_{\text{F}}|=3$ look
to be more simple. Assuming that $|m_{\text{F}}|=3$ resonance is a
simple resonance we try to understand the line shape with a simple
model.

\subsection{Theoretical approach: model for S-matrix.}

In order to describe the measured resonances, we need to account
for the inelastic collision processes. Feshbach resonances in cold
atomic systems are usually considered only in the context of
elastic scattering, with one open channel
only~\cite{moerdijk,marcelis}. In such a situation, only an
elastic resonance width $\Gamma_e$ is involved, and the incoming
channel for the atoms is simultaneously the exit channel. Here we
consider a situation with more than one open channel. Atoms which
exit in a different channel than the incoming channel will gain
additional kinetic energy, and are considered as lost. This decay
process can be mediated via the molecular Feshbach state, and
gives rise to an additional energy scale, the inelastic resonance
width $\Gamma_i$ \cite{feshbach}.

The S-matrix for such an inelastic resonance can be written as

\begin{equation}\label{eq:smat}
S(k)=S^{bg}(k) \left(
1-\frac{i\Gamma_e(k)}{E-\nu+\frac{1}{2}i\Gamma_e(k)+\frac{1}{2}i\Gamma_i(k)}
     \right)
\end{equation}
with $E=\hbar^2k^2/m$ the relative kinetic energy between the
colliding atoms, $\nu=\Delta \mu (B-B_0)$ the detuning, and
$S^{bg}(k)=\exp(-2ika_{bg})$ the background (or direct) part of
the S-matrix. Here $B_0$ is the magnetic field value of resonance,
and $\Delta\mu$ the difference between the magnetic moments of the
two particles in the incoming channel and the molecular Feshbach
state. The background scattering length
$a_{bg}=a_{bg}^r+ia_{bg}^i$ contains not only a real part
$a_{bg}^r$, but also an imaginary part $a_{bg}^i$ since there is
also a direct contribution to the inelastic decay.

The energy widths are in principle a function of the relative
wavenumber $k$, since they depend on the overlap between the
molecular wavefunction corresponding to the Feshbach state, and
the wavefunction in the corresponding open channel. In case of the
elastic energy width, the energy dependence of the incoming
channel gives rise to linear dependence in wavenumber, i.e.
$\Gamma_e=c_e k$ with $c_e$ a constant~\cite{moerdijk}. For the
inelastic energy width, however, the energy dependence of the
corresponding open channels can be safely neglected over the
energy range of interest. Therefore $\Gamma_i$ can be taken
energy-independent.

\subsection{Monte Carlo model.}

The frequency shift of the clock transition due to the $|3;m_F\neq
0\rangle$ state population is given by Eq.~(\ref{eq:clockshift}):
    \begin{equation}\label{eq:clockshift}
    \frac{\delta\omega_{\beta\alpha}}{2\pi}=\frac{\hbar \rho_{\gamma\gamma}}{m
    k}Im\left\{S_{\alpha\gamma}(k)S^{\dag}_{\beta\gamma}(k)-1\right\}
    \end{equation}
where $\alpha$ and $\beta$ refer to the $|3;0\rangle$ and
$|4;0\rangle$ clock states respectively. $\rho_{\gamma\gamma}$
    is the atomic density of the $|3;m_{\text{F}}\neq 0\rangle$ state (denoted as $\gamma$). To be consistent with our initial intent to analyze the
    $|m_{\text{F}}|=3$ data as a single Feshbach resonance, we will assume that a
    resonance occurs for only one of the two entrance channels ($\alpha\gamma$ and
    $\beta\gamma$). The other entrance channel contribute only with
    a constant background part, independent on the magnetic field. This factor can
    be formally accounted for in the background term of the channel where the resonance
    occurs. Therefore, we are assuming that $S_{\alpha\gamma}=1$ (resp.
    $S_{\beta\gamma}=1$) while $S_{\beta\gamma}(k)$ (resp.
    $S_{\alpha\gamma}$(k)) follows Eq.~(\ref{eq:smat}).

    Eq.~(\ref{eq:clockshift}) clearly shows that the clock
    shift depends on the wavenumber $k$. The total clock shift is
    obtained by suitably averaging Eq.~(\ref{eq:clockshift}) over the
    duration of the interrogation, taking into account the atomic
    cloud space and velocity distribution. This is done
    using a Monte Carlo simulation where the atomic trajectories
    are randomly sampled according to the space and
    velocity distributions as measured in the experiment (with CCD imaging and analysis of time of flight
    signals). The simulation accounts for the distribution of
    collisional energy, for the time varying (decreasing) atomic
    density, for the non-uniform response of the final transition probability
    to a time varying perturbation (sensitivity function),
    for the truncation of the atomic cloud crossing the microwave
    resonator.

\begin{figure}
  \includegraphics[width=8cm]{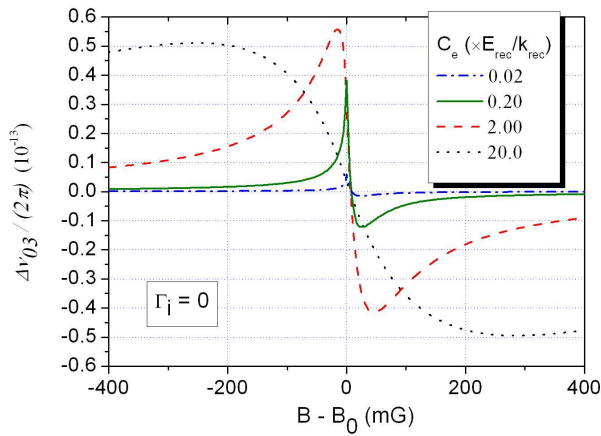}\\
  \caption{ For $\Gamma_i=0$. For large values of
    $C_e$, the shape and the (``natural") width of the resonance is determined
    by $C_e$. The resonance has a symmetrical dispersive shape.
    For small values of $C_e$, the shape and the width of the
    resonance is determined by the collision energy distribution.
    The line shape shows a strong asymmetry. The calculation
    clearly indicates that the top of the sharp feature corresponds
    to the magnetic field that meets the resonant condition.}\label{fig:model}
\end{figure}

\begin{figure}
  \includegraphics[width=8cm]{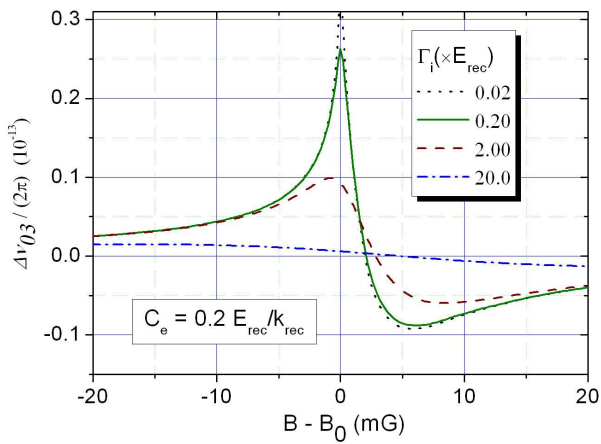}\\
  \caption{illustrates the effect of parameter
    $\Gamma_i$. $C_e$ is kept constant. The resonance shape is computed for different $\Gamma_i$ amplitudes. The comparison with the
    measured line shape clearly indicates that the observed
    resonance are not strongly affected by the inelastic processes.
    The calculation shows that the sign of the clock shift is inverted
    when the resonance occur on the $\alpha\gamma$ channel.}\label{fig:diversgammai}
\end{figure}

    FIGs.~\ref{fig:model} and \ref{fig:diversgammai} represent examples of calculated line shape of the Feshbach
    resonance for various parameters. For all cases, we take
    $a_{bg}=0$, $B_0=0$ and $\Delta\mu=2\mu_B$. Also, we assume that the resonance
    occurs on the $\beta\gamma$ channel. The initial distribution is gaussian with
    $\sigma=3.5$~mm. The velocity distribution is
    $\propto(1+v^2/v_c^2)^{-b}$ with $v_c=10$~mm.s$^{-1}$ and
    $b=2.1$. The vertical axis represents the fractional frequency shift due to the
    $\gamma$ ($|3;m_F\rangle$) state for $10^8$ atoms launched in this
    state. This corresponds to a typical effective atomic density
    of $2.2~10^7$~cm$^{-3}$.
    In FIG.~\ref{fig:model}, $\Gamma_i=0$. $C_e$ is equal to
    $0.2\times E_{rec}/k_{rec}$, $2\times E_{rec}/k_{rec}$,
    $20\times E_{rec}/k_{rec}$ and $200\times E_{rec}/k_{rec}$ where
    $E_{rec}=\hbar^2 k_{rec}^2/2m$ and
    $k_{rec}=2\pi/\lambda,~\lambda=852$~nm. For large values of
    $C_e$, the shape and the (``natural") width of the resonance is determined
    by $C_e$. The resonance has a symmetrical dispersive shape.
    For small values of $C_e$, the shape and the width of the
    resonance is determined by the collision energy distribution.
    The line shape shows a strong asymmetry. The calculation
    clearly indicates that the top of the sharp feature corresponds
    to the magnetic field that meets the resonant condition (vanishing difference
    between the molecular bound state energy and the dissociation threshold).
    Similarly, FIG.~\ref{fig:diversgammai} illustrates the effect of parameter
    $\Gamma_i$. $C_e$ is kept constant and equal to $0.2\times
    E_{rec}/k_{rec}$. The resonance shape is computed for
    $\Gamma_i$ equal to $0.02\times E_{rec}$, $0.2\times E_{rec}$, $2\times
    E_{rec}$ and $20\times E_{rec}$. The comparison with the
    measured line shape clearly indicates that the observed
    resonance are not strongly affected by the inelastic processes.
    Finally, the calculation (as well as direct consideration of Eq.~(\ref{eq:smat}) shows that the sign of the clock shift is inverted
    when the resonance occur on the $\alpha\gamma$ channel.

\begin{figure}
  \includegraphics[width=8cm]{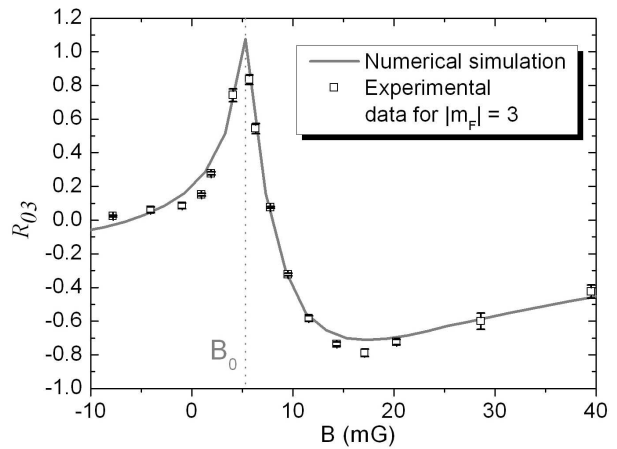}\\
  \caption{The $|m_{\text{F}}|=3$ data are in good agreement with a
  resonance dominated by the collision energy. The Monte Carlo model fits well
   the point and indicates the resonant field.}\label{fig:reso3fit}
\end{figure}

    The measured $|m_{\text{F}}|=3$ resonance is fitted (see FIG.~\ref{fig:reso3fit}) using the above
    model leaving $B_0$, $\Delta\mu$, $C_e$, $\Gamma_i$, $a_{bg}$ as free
    parameters.
    The model also includes the possibility of a slow variation of the
    background
    scattering length with the magnetic field through the addition of a linear and quadratic term as a function of the
    magnetic field. We checked that the following analysis is not significantly
    affected by this additional term. Finally, the model includes a scale factor for the atomic
    density. Similarly, the inclusion of this scale factor has no influence on the following analysis.

    The simple consideration of the sign of the data indicates that the resonance occurs in the $\beta\gamma$
    channel involving the upper clock state $|4;0\rangle$ and the $|3;3\rangle$ state. FIG.~\ref{fig:reso3fit} shows the optimized fit to the
    data. The qualitative agreement with the model is
    quite satisfactory. For $\Delta\mu$, $C_e$ and $\Gamma_i$, the
    analysis does not lead to precise value, due to the
    sensitivity of these parameters to the atomic cloud space and
    velocity distribution. We find $\Delta\mu\simeq 1.5 \mu_B$, $C_e \sim 0.2
    E_{rec}/k_{rec}$ and $\Gamma_i \lesssim 0.2\times E_{rec}$.
    Conversely, $B_0$ is tightly constrained and well
    decorrelated from the other parameters. Although an uncertainty on $B_0$ cannot be easily extracted from our (non-linear) fitting procedure,
    we can safely state that $B_0$ is constrained to
    within $1$~mG (to be understood as a $1\sigma$ error bar). We
    find $B_0=5\pm 1$~mG, this corresponds to $500\pm100$~nK, expressed in temperature scale. This is, to our knowledge,
     the lower molecular bound state energy ever involved in Feshbach resonances (\textit{i.e.} around 10~kHz from the continuum).

\section{Conclusion.}

To sum-up, our primary goal is achieved, the collisional shift is
controlled at the $10^{-3}$ of its value. A calibration of Zeeman
sub-states contribution to the clock shift as a function of the
field has been performed and teaches us magnetic field values to
avoid. Under regular clock conditions an upper limit for their
contribution shows an upper limit compatible with AP at the
$10^{-3}$ level. This makes reachable the goal of primary
standards at the $10^{-16}$ uncertainty level. Feshbach resonances
have been observed for the first time at very low magnetic field
and with a very good resolution. A Monte Carlo simulation has been
performed and could fit properly some of experimental
data. This constrains some parameters of the theory of collisions.\\

\noindent \textit{Acknowledgments}: The authors wish to thank K.
Williams for fruitful discussions. This work was supported in part
by BNM and CNRS. BNM-SYRTE and Laboratoire Kastler-Brossel are
Unit\'es Associ\'ees au CNRS, UMR 8630 and 8552. S.K.~acknowledges
supported from the Netherlands Organization for Scientific
Research (NWO).


\begin{thebibliography}{30}
\bibitem{Anderson95} M.H. Anderson {\it et al.}, Science {\bf 269}, 198 (1995).
\bibitem{Bijlsma94} M. Bijlsma \textit{et al}., Phys. Rev. A {\bf 49}, R4285 (1994).
\bibitem{Bizeieee01} S. Bize \textit{et al}., IEEE Trans. on Instr. and Meas. {\bf 50}, 503 (2001)
\bibitem{Bize01} S. Bize \textit{et al}., in \textit{Proc. of the 6$^{th}$ Symposium on Frequency Standards and Metrology} (World Scientific, Singapore, 2001), p 53.
\bibitem{Bize03} S. Bize \textit{et al}., Phys. Rev. Lett. \textbf{90}, 150802 (2003).
\bibitem{chambon04} D. Chambon \textit{et al}., Proceeding EFTF 2004.
\bibitem{Chin01} C. Chin \textit{et al}., Phys. Rev. A {\bf 63}, 033401 (2001).
\bibitem{Feshbach} H. Feshbach, Ann. of Phys. (N.Y.) \textbf{5}, 357 (1958); \textbf{19}, 287 (1962).
\bibitem{feshbach} H. Feshbach, \textit{Theoretical Nuclear Physics} (John Wiley and Sons, New York, 1992).
\bibitem{Ghezali96} S. Ghezali \textit{et al}., Europhys. Lett. {\bf 36}, 25 (1996).
\bibitem{Gibble93} K. Gibble and S. Chu, Phys. Rev. Lett. {\bf 70}, 1771 (1993).
\bibitem{Loy74} M.M.T. Loy, Phys. Rev. Lett. {\bf 32}, 814 (1974).
\bibitem{luiten95} A. Luiten {\it et al.}, IEEE Trans. on Instr. and Meas. {\bf 44}, 132 (1995).
\bibitem{marcelis}  B. Marcelis, E.G.M. van Kempen, B.J. Verhaar, and S.J.J.M.F. Kokkelmans, cond-mat/0402278.
\bibitem{Marion03} H. Marion {\it et al.}, Phys. Rev. Lett. {\bf 90}, 150801 (2003).
\bibitem{TheseHarold} H. Marion, \textit{Controle des collisions froides du 133Cs, test de la variation de la constante de structure fine à l'aide d'une
    fontaine atomique double rubidium - cesium}, PhD Thesis, Universite Paris VI, (2005).
\bibitem{Messiah} A. Messiah, Quantum Mechanics, {\bf 2},637 (1959).
\bibitem{moerdijk} A.J. Moerdijk, B.J. Verhaar, and A. Axelsson, Phys.~Rev.~A {\bf 51}, 4852 (1995).
\bibitem{Pereira02} F. Pereira Dos Santos {\it et al.}, Phys. Rev. Lett. {\bf 89}, 233004 (2002).
\bibitem{rasel99} E. Rasel \textit{et al}., Eur. Phys. J. D {\bf 7}, 311-316 (1999).
\bibitem{Santarelli99} G. Santarelli {\it et al.}, Phys. Rev. Lett. {\bf 82}, 4619 (1999)
\bibitem{shimizu90} F. Shimizu \textit{et al}., Chem. Phys. {\bf 145}, 327 (1990).
\bibitem{Sortais00} Y. Sortais {\it et al.}, Phys. Rev. Lett.  {\bf 85}, 3117 (2000).
\bibitem{Tiesinga99} E. Tiesinga {\it et al.}, Phys. Rev. A \textbf{47}, 4114 (1993).
\bibitem{Wolf03} P. wolf {\it et al.}, Phys. Rev. Lett. {\bf90}, 060402 (2003).
\bibitem{Wynands} R.Wynands and S. Weyers, Atomic fountain clocks, Metrologia {\bf42} S64–S79, (2005)




\end{thebibliography}
\end{document}